\documentclass[11pt]{article}
\usepackage{amssymb}
\usepackage{citesort}
\usepackage{deleq}
\usepackage{url}
\usepackage{a4}
\usepackage{cite}
\usepackage{ntheorem}
\newcommand{\mcNpr}{\mcN^+_r}
\newcommand{\mcNpl}{\mcN^+_l}
\newcommand{\mcNmr}{\mcN^-_r}
\newcommand{\mcNml}{\mcN^-_l}
\newcommand{\myphi}{\phi}
\newcommand{\myh}{\gamma}

\newcommand{\Ydown}{Y^\flat}

\newcommand{\riemgz}{g_0}

\renewcommand{\hbar}{{\overline \riemgz}}

\newcommand{\nablash}{\nabla{\kern -.75 em
     \raise 1.5 true pt\hbox{{\bf/}}}\kern +.1 em}
\newcommand{\Deltash}{\Delta{\kern -.69 em
     \raise .2 true pt\hbox{{\bf/}}}\kern +.1 em}
\newcommand{\Rslash}{R{\kern -.60 em
     \raise 1.5 true pt\hbox{{\bf/}}}\kern +.1 em}

\newcommand{\Ric}{\operatorname{Ric}}

\newcommand{\mcO}{{\mycal O}}

\newcommand{\hyp}{\Sigma}

\newcommand{\mcM}{{\mycal M}}
\newcommand{\mcD}{{\mycal D}}
\newcommand{\mcN}{{\mycal N}}
\newcommand{\mcS}{{\cal S}}
\newcommand{\mcK}{{\mycal K}}

\newcommand{\bea}{\begin{eqnarray}}
\newcommand{\beaa}{\begin{eqnarray*}}
\newcommand{\bean}{\begin{eqnarray}\nonumber}

\newcommand{\bel}[1]{\begin{equation}\label{#1}}
\newcommand{\beal}[1]{\begin{eqnarray}\label{#1}}
\newcommand{\beadl}[1]{\begin{deqarr}\label{#1}}
\newcommand{\eeadl}[1]{\arrlabel{#1}\end{deqarr}}
\newcommand{\eeal}[1]{\label{#1}\end{eqnarray}}
\newcommand{\eead}[1]{\end{deqarr}}
\newcommand{\eea}{\end{eqnarray}}
\newcommand{\eeaa}{\end{eqnarray*}}

\newcommand{\be}{\begin{equation}}
\newcommand{\ee}{\end{equation}}

\newcommand{\eq}[1]{\eqref{#1}}
\newcommand{\Eq}[1]{Equation~(\ref{#1})}

\DeclareFontFamily{OT1}{rsfs}{}
\DeclareFontShape{OT1}{rsfs}{m}{n}{ <-7> rsfs5 <7-10> rsfs7 <10->
rsfs10}{} \DeclareMathAlphabet{\mycal}{OT1}{rsfs}{m}{n}

\usepackage{amsmath} 

\renewcommand{\theequation}{\thesection.\arabic{equation}}

\let\ssection=\section

\renewcommand{\section}{\setcounter{equation}{0}\ssection}

\newtheorem{defi}{\sc Definition\rm}[section]

\newtheorem{Proposition}[defi]{\sc Proposition\rm}

\theorembodyfont{\upshape}

\def \Reel{\mathbb{R}}
\def \C{\mathbb{C}}
\def \R {\Reel}

\newcounter{mnotecount}[section]

\newcommand{\rmnote}[1]{}

\begin{document}
\title{On analyticity of static vacuum metrics at non-degenerate horizons}
\author{
Piotr T. Chru\'sciel\thanks{Partially supported by a Polish
Research Committee grant 2 P03B 073 24; email \protect\url{
piotr@gargan.math.univ-tours.fr}, URL \protect\url{
www.phys.univ-tours.fr\~piotr}}
\\
 D\'epartement de math\'ematiques\\ Facult\'e des
Sciences\\ Parc de Grandmont\\ F37200 Tours, France}
\date{}

\maketitle


\begin{abstract}
We show that static metrics solving vacuum Einstein equations
(possibly with a cosmological constant) are one-sided analytic at
non-degenerate Killing horizons. We further prove analyticity in a
two-sided neighborhood of ``bifurcate horizons".

{\sl Cracow's school of physics, led by Prof. Staruszkiewicz, has
made deep contributions to our understanding of black holes. It is
a pleasure to dedicate to Prof. Staruszkiewicz this contribution
to the subject, on the occasion of his 65th birthday.}
\end{abstract}






\section{Introduction}
It is a classical result of M\"uller zum Hagen~\cite{MzH} that
stationary vacuum metrics are analytic, in appropriate charts, in
the region where the Killing vector is timelike. However,
analyticity does sometimes stop at Killing horizons, as can be
seen by the Scott-Szekeres extensions of the Curzon
metric~\cite{ScottSzekeresI,ScottSzekeresII};
compare~\cite{Bicak:podolsky} for examples with a cosmological
constant. The aim of this note is to point out that one-sided
analyticity always holds at \emph{non-degenerate} static Killing
horizons. We also prove analyticity in a (two-sided) neighborhood
of ``bifurcate horizons". In addition to their intrinsic interest,
our results have applications to the classification of static
solutions\footnote{In his proof of Israel's theorem,
Robinson~\cite{RobinsonSP} appeals to analyticity up-to-boundary
of the metric, which has not been justified until this work. While
Robinson's proof has been superseded by more complete
results~\cite{bunting:masood,Chstatic}, it remains the simplest
one in the connected non-degenerate case, and it seems of interest
to have a complete argument along his lines.} of the Einstein
equations with $\Lambda=0$~\cite{RobinsonSP} or with
$\Lambda>0$~\cite{BessieresLafontaineRozoy}, or to discussions of
cosmic censorship~\cite{VinceJimcompactCauchyCMP}
(compare~\cite{FRW}).

We assume an arbitrary space-time dimension $n+1$, and vacuum
Einstein equations, with or without a cosmological constant. It
should be clear that the argument generalises to certain couplings
of matter fields to the geometry via  Einstein equations.

We expect the result to remain valid for stationary, not
necessarily static, Killing horizons, we plan to return to this
question in a near future.

\section{The method}\label{Smethod}
The proof of the above turns out to be rather simple, and relies
on the ``Wick rotation" method.  It is well known that  a metric
$g$ with timelike Killing vector field $X$ may locally be written
in the form
\begin{equation} \label{e1.2}
g = -u^{2}(dt  + \theta_idx^i )^{2} + h_{ij}dx^i dx^j\;,
\end{equation}
where $\theta_idx^i $ is a connection 1-form on the space of
$x^i$'s, $u$ is the length of the Killing field $X = \partial
/\partial t$, and $h$ is a Riemannian metric. All the fields above
are $t$-independent. (Since all considerations here are strictly
local the range of the function $t $, and  the associated question
of completeness of the orbits of $X$, are completely irrelevant
for our purposes.) The simplest case to consider is that of static
metrics, where $\theta$ can be set to zero, so that \eq{e1.2}
becomes
\begin{equation} \label{e1.2n}
g = -u^{2}dt^{2} + h_{ij}dx^i dx^j\;.
\end{equation}
Suppose that $g$ solves the vacuum Einstein equations (possibly
with a cosmological constant), it is well known that the
Riemannian counterpart of $g$, \bel{e1.21} u^{2}d\tau^{2} +
h_{ij}dx^i dx^j\;,\ee also satisfies those equations. A simple way
of seeing that is as follows: for $\alpha\in \C^{*}$ consider the
family of complex valued tensor fields
$$g(\alpha)= -\alpha^2 u^{2}dt^{2} + h_{ij}dx^i dx^j\;.$$
Let $\Ric(\alpha)$ be the complex valued tensor field obtained by
calculating the Ricci tensor of $g(\alpha)$ using the usual
formulae. Since the Ricci tensor is a rational function of the
$g_{\mu\nu}$'s and their derivatives, all the coordinate
components $R(\alpha)_{\mu\nu}$ of $\Ric(\alpha)$ are meromorphic
functions of $\alpha$. For $\alpha\in \R^*$ we have
$R(\alpha)_{\mu\nu}=0$, since for those values the metric
$g(\alpha)$ can be obtained by a coordinate transformation
$t\to\tau= \alpha t $ from the metric $g=g(1)$. Uniqueness of
analytic extensions implies that $R(\alpha)_{\mu\nu}=0$ for all
$\alpha\in\C^*$, setting $\alpha=i$ one obtains the desired result
for the Riemannian metric $g(i)$.

An identical argument applies of course to the family of complex
tensor fields
\begin{equation} \label{e1.3}
g(\alpha) = -u^{2}(\alpha dt  + \theta_jdx^j )^{2} + h_{jk}dx^j
dx^k\;,
\end{equation}
so that if $g$ were an Einstein Lorentzian metric,
\bel{e1.3c}\Ric(g) = \lambda g\ee for some constant $\lambda$, then
the complex tensor field $g(\alpha)$ again satisfies the set of
equations
\bel{e1.3a}\Ric\left(g(\alpha)\right)=\lambda g(\alpha)\ee for all values of $\alpha\in
\C^*$. In particular if $\alpha=i$ we obtain that the complex
tensor field
\begin{equation} \label{e1.3b}
g(i) = u^{2}( d\tau  + i\theta_jdx^j )^{2} + h_{jk}dx^j dx^k\;,
\end{equation}
solves the set of complex equations \eq{e1.3c}. In this work we
will, however, concentrate on the static case, so that this fact
is irrelevant for the remainder of this paper.

\section{One-sided analyticity near a Killing horizon}\label{Secsst}
From now on we restrict ourselves to the static case, locally
$\theta=df$. Recall that a Killing horizon is a null hypersurface
$\mcK$ such that $X$ is tangent to the generators of $\mcK$. As is
well known (see, \emph{e.g.},~\cite[Proposition~3.2]{Chstatic}), a
non-degenerate $\mcK$ corresponds to a smooth totally geodesic
boundary, say $\partial \hyp$, for the metric $h$. Further, if
$\rho=\rho(p)$ denotes the distance from $p$ to $\partial \hyp$ in
the metric $h$ then, in Gauss coordinates around $\partial \hyp$,
all the functions appearing in the metric are
smooth\footnote{\label{Fdif}Throughout we assume smoothness of the
manifold and of the metric. However, there exists $k<\infty$ such
that if the metric is $C^k$, then the methods here apply, leading
to analyticity. The exact value of $k$ can be found by chasing
losses of differentiability that arise in the constructions here,
as well as in those of~\cite{Chstatic}.} functions of $\rho^2$ and
of the remaining coordinates. Moreover, $u$ vanishes on $\partial
\hyp=\{\rho=0\}$, with non-zero gradient there. This implies (the
well known fact) that the set $\{\rho=0\}$ for the Riemannian
metric \eq{e1.21} corresponds to a smooth axis\footnote{By this we
mean a submanifold of codimension two invariant under the flow of
$Y$, with $Y$ generating rotations in the normal bundle.} of
rotation of a Killing vector $Y$. Now $Y$ is the obvious
counterpart of $X$ under the transition from \eq{e1.2n} to
\eq{e1.21}, and this transition preserves
hypersurface-orthogonality, hence $Y$ satisfies $$\Ydown \wedge
d\Ydown = 0\;,$$ where $\Ydown:= g(i)(Y,\cdot)$. Since $g(i)$ is a
Riemannian Einstein metric, its coordinate components $g(i)_{ij}$,
with respect to harmonic coordinates, satisfy an elliptic
quasilinear system of PDEs and are, therefore, real analytic.
Further, the harmonic coordinates are smooth in the original
smooth atlas. The geodesic coordinates around the rotation axis
$\partial \Sigma$ are also analytic because 1) the axis of
rotation is a totally geodesic submanifold (of co-dimension two)
in the Riemannian manifold $(M,g(i))$, hence analytic; 2) normal
coordinates around an analytic submanifold are analytic in an
atlas in which the metric is analytic. (This follows from the
analytic implicit function theorem~\cite{Shabat}.) It should be
clear that this provides the desired \emph{one-sided} analytic
atlas in the Lorentzian solution near the horizon, by running
backwards the calculations of,
\emph{e.g.},~\cite[Proposition~3.2]{Chstatic}. Since there is a
major subtlety here, as
$$\mbox{\em one obtains analyticity only in the region
$g(X,X)\le0$},$$ we provide the details: Consider a covering of
$\{\rho=0\}$ by domains of definition $\mcO_i$, $i=1,\cdots, N$,
of analytic coordinate systems $x^a$, $a=3,\ldots,n+1$, and for
$q\in \mcO_a$ let $x^A$, $A=1,2$, denote geodesic coordinates on
$\exp_q\{(T_q\partial \Sigma)^\perp\}$. Set $(x^\mu)=( x^A,x^a)$.
From what has been said it follows that the $x^\mu$--coordinate
components of the Riemannian metric tensor $g(i)$ are analytic
functions of the $x^\mu$'s.
 We have  the following local form of the metric
\bel{afp0} g(i)= \sum_{i=1}^{2}
(dx^i)^2 + h +\sum_{A,a}O(\rho)dx^Adx^a +
\sum_{A,B}O(\rho^2)dx^Adx^B +\sum_{a,b}O(\rho^2)dx^adx^b   \;,\ee
with $h$
--- the metric induced by $g$ on $\partial \Sigma$. The $O(\rho^2)$ character of the $dx^A
dx^B$ error terms is standard;  the $O(\rho^2)$ character of the
$dx^a dx^b$ error terms follows from the totally geodesic
character of $\partial\Sigma$. The Killing vector field $Y$ takes
the form $Y=x^1\partial_2 - x^2\partial_1=\partial_\varphi$, where
\bel{polar}(x^1,x^2)=(\rho \cos \varphi, \rho \sin \varphi)\;.\ee When
expressed in terms of $\rho$ and $\varphi$, the functions
$g(i)_{\mu\nu}:=g(i)(\partial_{x^\mu},\partial_{ x^\nu})$ are
analytic functions of the $x^\mu$'s, hence (by composition) of
$\rho$ and
 of $\varphi$. Let $R_\pi$ denote a rotation by $\pi$ in the
$(x^A)$-planes, $R_\pi$ is obtained by flowing along $Y$ a
parameter-time $\pi$ and is therefore an isometry, leading to
\beadl{par1} & g(i)_{ab}(-x^1,-x^2,x^a)= g(i)_{ab}(x^1,x^2,x^a)\;,
&\\& g(i)_{AB}(-x^1,-x^2,x^a)= g(i)_{AB}(x^1,x^2,x^a)\;,&\\&
g(i)_{Aa}(-x^1,-x^2,x^a)= -g(i)_{Aa}(x^1,x^2,x^a)\;.\eeadl{par} In
particular all odd-order derivatives of $g_{ab}$ with respect to
the $x^B$'s vanish at $\{x^A=0\}$, \emph{etc.} Those symmetry
properties together with analyticity imply (using, e.g., Osgood's
lemma) that there exist analytic $b_{ab}(s,x^a)$,
$\gamma_a(s,x^a)$, $\psi(s,x^a)$, with $\psi(0,x^a)=1$, such that
$$g(i)_{ab}(x^1,x^2,x^a)= b_{ab}(\rho^2,x^a)\;,$$
$$ \left(g(i)_{Ab}Y^A\right)(x^1,x^2,x^a)= \rho^2 \gamma_{b}(\rho^2,x^a)\;.$$
$$u(x^1,x^2,x^a):=\sqrt{\left(g(i)(Y,Y)\right)(x^1,x^2,x^a)}= \rho\left(1+\rho^2 \psi(\rho^2,x^a)\right)\;.$$
Similarly, let $n=x^A\partial_A$, then $g(i)_{AB}Y^A n^B $ and
$g(i)_{AB}n^A n^B$ are analytic functions invariant under the flow
of $Y$, with $g(i)_{AB}Y^A n^B=(g(i)_{AB}-\delta_{AB})Y^A n^B=
O(\rho^4)$, $g(i)_{AB}n^A n^B=\rho^2+O(\rho^4)$, hence there exist
analytic functions $\alpha(s,x^a)$ and $\beta(s,x^a)$ such that
$$\left(g(i)_{AB}Y^A n^B\right)(x^1,x^2,x^a)= \rho^4 \alpha(\rho^2,x^a)\;,
$$ $$ \left(g(i)_{AB}n^A n^B\right)(x^1,x^2,x^a)=\rho^2 + \rho^4
\beta(\rho^2,x^a)\;.$$
One similarly finds existence of an analytic one-form
$\lambda_a(s,x^b)dx^a$ such that
$$\left(g(i)_{Aa}n^A \right)(x^1,x^2,x^b)= \rho^2
\lambda_a(\rho^2,x^b)\;.$$ In polar coordinates \eq{polar} one
therefore obtains
$$\Ydown :=g(i)(Y,\cdot)= \rho^2\left( (1+\rho^2\psi)^2 d\varphi + \alpha \rho d\rho +
\gamma_a dx^a\right)\;.$$ Writing $g(i)$ in the form\footnote{Note
that $\theta_j$ here is real, arising from the potential failure
of hypersurface orthogonality of the polar coordinates associated
to the harmonic ones, and {\em not} from the introduction of a
complex constant in the metric as in Section~\ref{Smethod}. The
introduction of the complex constant $i$ there was done only to
justify that the Riemannian metric denoted by $g(i)$ is Einstein;
this last fact can be checked by direct calculations in any case.}
\begin{equation} \label{e1.3bn}
g(i) = u^{2}( d\varphi  + \theta_jdy^j )^{2} + h_{jk}dy^j dy^k\;,
\end{equation}
with $y^j=(\rho^2, x^a)$, one has $\Ydown =u^2(d\varphi
+\theta_jdy^j)$ leading to
$$\theta:=\theta_jdy^j = \frac {\alpha}{2(1+\rho^2\psi)^2}d(\rho^2) +
(1+\rho^2\psi)^{-2}\gamma_adx^a\;,$$ \bel{hred}h_{jk}dy^j dy^k=
(1+\rho^2\beta) \left(\frac{d(\rho^2)}{2\rho}\right)^2 +
b_{ab}dx^a dx^b + \lambda_a d(\rho^2)dx^a- u^2 \theta_i \theta_j
dy^i dy^j \;, \ee in particular all the functions $h_{jk}$ are
analytic functions of $\rho^2$ and $x^a$, except for the singular
term $(2\rho)^{-2}\left(d(\rho^2)\right)^2$.

Note that hypersurface-orthogonality has not been used anywhere in
the calculations above (except for the initial justification of
analyticity),\footnote{It should be emphasised that, from a
space-time point of view, the hypothesis of non-degeneracy of the
horizon has been made. We are not aware of any results about the
behavior of $\theta$ near degenerate horizons.} so that quite
generally we have proved:

\begin{Proposition}
\label{Lnohg} $\theta$ extends smoothly to the rotation axis
$\{Y=0\}$, analytically when the metric is analytic.
\end{Proposition}

Let us return to the static case, the hypersurface-orthogonality
condition $\Ydown \wedge d\Ydown =0$ is equivalent to $d\theta=0$,
hence there exists, locally, a function $\tau $ such that
$$ d\tau=d\varphi + \theta\;.$$
(The function $\tau$ is clearly analytic in the $y^i$'s, but this
is irrelevant for our purposes, since all the functions in
\eq{e1.3bn} are $\varphi$-independent.) Writing \bel{rema}
h_{jk}dy^j dy^k= \left(\frac{d(\rho^2)}{2\rho}\right)^2 + \hat
h_{jk}dy^j dy^k\;,\ee where the $\hat h_{ij}$'s are defined by
subtracting the first term at the right-hand-side of \eq{rema}
from \eq{hred}, the Lorentzian equivalent of the metric
\eq{e1.3bn} reads now
\bel{finm} g=g(1)= -y^1(1+y^1\psi)^2 dt^2 +\frac{(dy^1)^2}{4y^1} +
\hat h_{jk}dy^j dy^k\;.\ee Introducing a new coordinate $u$
replacing $t$, $$ u=t+\frac 12 \ln (y^1)\;,$$ the undesirable
singular term in \eq{finm} cancels out. This provides the required
analytic atlas in a one-sided neighborhood of the Killing horizon,
covering the set $\{g(X,X)\le 0\;,\ X\ne 0\}$, compatible with the
initial smooth structure, in which the metric functions are
analytic up-to-boundary on the set  where $X$ is timelike or null.

\section{Static initial data: global analyticity}
\label{Sstaticid} In the previous section  the starting point of
our considerations was a static space-time. However, one can start
with static initial data and ask about regularity of those. More
precisely, consider a triple $(M,\myh ,\myphi )$, where $(M,\myh)$
is a smooth$^{\mbox{\rm \scriptsize \ref{Fdif}}}$ $n$-dimensional
Riemannian manifold and $\myphi $ is a smooth function on $M$,
satisfying the following set of equations
\beadl{sve1}&\Delta_\myh \myphi  = - \lambda \myphi \;,&
\\ &\myphi (R(\myh)_{ij}-\lambda \myh_{ij}) = D_iD_j \myphi \;.&\label{sve2}
\eeadl{sve} Here $D$ is the Levi-Civita connection of $\myh$,
$\Delta_\myh:=D_kD^k$ its Laplace-Beltrami operator,
$R(\myh)_{ij}$ the Ricci tensor of $\myh$, $R(\myh)$ the scalar
curvature of $\myh$, while $\lambda \in \R$ is a constant related
to the cosmological constant $\Lambda$. The function $\myphi $ is
allowed to change sign. It is well known that the set of zeros of
a non-trivial $\myphi $, solution of \eq{sve}, forms a smooth,
embedded, totally geodesic  submanifold of $M$, if not empty.
Again, it is well known~\cite{MzH} that $M\setminus \{\myphi =0\}$
can be endowed with an analytic atlas, with respect to which
$\myh$ and $\myphi $ are analytic --- this is a relatively
straightforward consequence of the underlying elliptic features of
the system of equations \eq{sve} for $\myh$ and $\myphi $, in
regions where $\myphi $ does not contain zeros.

We wish to show analyticity up-to and across the set of zeros of
$\myphi $: Consider, thus, a one-sided local neighborhood of
$\{\myphi =0\}$, replacing $\myphi $ by $-\myphi $ if necessary it
suffices to consider the case $\myphi \ge 0$. An appropriate
periodic identification of an angular variable $\phi $ shows that
the Riemannian metric, which we shall call $g(i)$,
$$g(i)= \myphi^2 d\varphi^2 + \gamma $$
has a smooth\footnote{This is established by first introducing
normal coordinates $(x,v^a)$ near $\{\myphi=0\}$, and using the
fact that $u=\kappa x + O(x^2)$, for some non-zero constant
$\kappa$. This provides continuity of the metric. To obtain
smoothness one can prove directly, using \eq{sve}, the parity
properties of $g(i)$ as in \eq{par1} with $x^2=0$, with a similar
equation for $\myphi/x$. Alternatively, it follows from \eq{sve}
that $R(\myh)=(n-1)\lambda$, therefore the set $(M,\myh,K:=0)$ is
a vacuum initial data set (with cosmological constant) for the
vacuum Einstein equations. Letting $(\mcM,g)$ be the maximal
globally hyperbolic development of the data, if $\{\myphi =0\}$ is
not empty then on $(\mcM,g)$ there exists a static
hypersurface-orthogonal Killing vector $X$ with a non-degenerate
Killing horizon, and \eq{par} follows from the analysis
in~\cite{Chstatic}.}  axis of rotation at $\{\varphi=0\}$ for a
Killing vector field $Y=\partial_\varphi$, and is Einstein.Then
the argument leading from \eq{afp0} to \eq{rema} applies, and is
actually somewhat simpler because in the Riemannian case there is
no need to introduce a new coordinate $y^1=\rho^2$, the coordinate
system $(\rho,x^a)$ being the one in which the metric is analytic.
\Eq{e1.3bn} shows that $h$ is the metric on the space of orbits of
the Killing vector $Y$, so is $\gamma$, hence $h$ is isometric to
$\gamma$. This proves one-sided analyticity of $\gamma$ in an
appropriate atlas. Similarly considering the region $\{\myphi\le
0\}$, one obtains an analytic atlas on $\{\myphi \le 0\}$ with
respect to which $-\myphi $ and $\myh$ are analytic. Thus, $\myphi
$ and $\myh$ are analytic up-to-boundary both on $\{\myphi \ge
0\}$ and on $\{\myphi  \le 0\}$. Smoothness implies that the power
series on both sides of $\{\myphi =0\}$ coincide, establishing
analyticity near $\{\myphi =0\}$, and hence throughout $M$.

\section{Analyticity in a neighborhood of a ``bifurcate Killing horizon"}\label{Secsstn}

 The one-sided analyticity of
the metric up-to-and-including the event horizon suffices for
several purposes; that is, \emph{e.g.}, the case for most issues
concerning the properties of static domains of outer
communications, which under the usual conditions coincide with the
set $g(X,X)<0$. It is, nevertheless, interesting to enquire about
extendibility of analyticity across the horizon. In the current
context the following examples should be borne in mind:
\begin{enumerate}
\item Consider any smooth vacuum space-time $(\mcM,g)$ with a non-degenerate Killing
horizon $\mcNpr $ associated to a Killing vector field $X$, and
suppose that $\mcM$ contains the ``bifurcation
surface"\bel{bifs}\mcS:=\{X=0\}\cap \overline{\mcNpr }\ne
\emptyset\;.\ee Let $\mcNpl  $ a second Killing horizon associated
with $\mcS$, so that $\dot J^+(\mcS)=\mcNpl  \cup \mcNpr $, see
Figure~\ref{F1}.
\begin{figure}\begin{center}\begin{center}
\begin{picture}(150,120)(0,-10)









 \thinlines
 \put(0,100){\line(1,-1){100}}

\put(0,0){\line(1,1){100}}




\put(50,50){\circle*{3}}


\put(75,25){\makebox(0,0)[l]{$\leftarrow\mcNmr$}}

\put(25,25){\makebox(0,0)[r]{$\mcNml\!\!\rightarrow$}}
\put(25,75){\makebox(0,0)[r]{$\mcNpl\!\!\rightarrow$}}


\put(52,50){\makebox(0,0)[l]{$\leftarrow\mcS\subset \{X=0\}$}}

%
\put(50,95){\makebox(0,0)[c]{$J^+(\mcS)$}}
%

\put(75,75){\makebox(0,0)[l]{$\leftarrow\mcNpr$}}

\end{picture}
\end{center}
\end{center}\caption{Four Killing
horizons $\mcN^\pm_r$ and $\mcN^\pm_l$ meeting at a bifurcation
surface $\mcS$. We have $J^+(\mcS)=\mcD^+(\mcNpl\cup\mcNpr)$, at
least locally.}\label{F1}
\end{figure}(In case of unusual global causality properties of $(\mcM,g)$,
the notions of future and past here should be understood locally
near $\mcS$.) Smoothly perturbing the characteristic initial data
on $\mcNpl  $, without modifying those on $\mcNpr $, by evolution
one will obtain a space-time $(\mcM',g')$ such that 1) $g'$
smoothly extends to the previous metric $g$ across $\mcNpr $; 2)
for generic perturbations there will be no Killing vectors on
$J^+(\mcNpl \cup \mcNpr )$. In the new space-time there will still
be a locally defined Killing vector field $X$ in a one-sided
(past) neighborhood of $\mcNpr $, but $X$ will not extend anymore
to a Killing vector field defined on $\mcM'$. Thus, even the
extendibility of a Killing vector field across a one-sided Killing
horizon might fail in general (compare, however,~\cite{FRW}). An
example of such behavior, with a metric which is explicit except
for one function, in the category of
 $C^{557}$ (but not smooth)
metrics, is provided by the family of Robinson-Trautman extensions
of the Schwarzschild metric of~\cite[Corollary~3.1]{ChS}.
\item Analyticity alone does not guarantee uniqueness of
extensions.
\end{enumerate}

In any case, so far we have only shown one-sided analyticity
up-to-boundary, on a set where $X$ is timelike or null, and it is
not completely clear that this will guarantee analyticity beyond
the Killing horizon in general: in each coordinate chart on which
the set $g(X,X)<0$ is given by $\{x^1>0\}$ there exists an
analytic extension of the metric to an open subset of the set
$\{x^1<0\}$, but this extension could fail to
coincide\footnote{Recall the example where the analytic
up-to-boundary function $f(x)=0$ defined for $x\ge 0$ is smoothly
extended by $\exp(x^{-1})$ for $x<0$.} with the original metric
there.

Let us show that there exists a setting where analyticity
necessarily extends beyond the event horizon: suppose, for
instance, that $\mcM$ contains a bifurcation surface $\mcS$ as in
\eq{bifs} (compare Figure~\ref{F1}) contained within a spacelike
achronal hypersurface $\hyp$. Assuming staticity, we can deform
$\Sigma$ in space-time so that $\Sigma$ is orthogonal to $X$. The
results in Section~\ref{Sstaticid} show that the initial data
induced on $\hyp$ are analytic with respect to an appropriate
atlas. By~\cite{AlinhacMetivier} the metric $g$ is analytic in
wave coordinates, compatible with the analytic atlas on $\hyp$, on
the domain of dependence $\mcD(\hyp)$. This last set contains a
neighborhood of $\mcS$. Let, now, $\mcNpr$ and $\mcNpl$ be as in
Figure~\ref{F1}. Section~\ref{Secsst} provides analytic
characteristic initial data (see, e.g.,~\cite{RendallCIVP}) there,
and we have already established analyticity in a whole space-time
neighborhood of $\mcS$. But analytic characteristic initial data
on $\mcNpr$ and $\mcNpl$, compatible at $\mcS$, lead\footnote{A
simple proof is obtained using Garabedian's
proof~\cite[Volume~III]{Taylor} of the Cauchy-Kowalevska theorem.}
to an analytic solution in $\mcD^+({\mcNpr \cup \mcNpl })$,
providing the desired result.

\bibliographystyle{amsplain}
\bibliography{
../references/newbiblio,%
../references/newbib,%
../references/reffile,%
../references/bibl,%
../references/Energy,%
../references/hip_bib,%
../references/netbiblio,../references/addon}

\end{document}